\documentclass[9pt,nofootinbib,superscriptaddress,longbibliography,twocolumn]{revtex4-1}

\usepackage{times}
\usepackage{graphicx}
\usepackage{dcolumn}
\usepackage{bm,bbm}
\usepackage{tabularx}
\usepackage{xcolor}
\usepackage{tcolorbox}
\usepackage{algorithm}
\usepackage{amsmath}
\usepackage{amsthm, amssymb}
\usepackage{makecell}
\usepackage{bbold}
\usepackage{braket}
\usepackage{nameref}
\usepackage[toc,page]{appendix}
\usepackage[colorlinks,breaklinks,linkcolor={blue},citecolor={magenta},urlcolor={blue}]{hyperref}

\usepackage{algorithmic}

\newtheorem{theorem}{Theorem}[section]
\newtheorem{prop}{Proposition}

\makeatletter
\newcommand\org@hypertarget{}
\let\org@hypertarget\hypertarget
\renewcommand\hypertarget[2]{%
  \Hy@raisedlink{\org@hypertarget{#1}{}}#2%
  }
\makeatother

\begin{document} 

\title{Optimal Quantum Circuit Design via Unitary Neural Networks}

\author{M. Zomorodi}
\affiliation{Department of Computer Science, Cracow University of Technology, Poland}

\author{H. Amini}
\affiliation{Department of Computer Engineering, Ferdowsi University of Mashhad, Iran}

\author{M. Abbaszadeh}
\affiliation{Department of Computer Science, University College London, London, UK}

\author{J. Sohrabi}
\affiliation{Department of Physics, Isfahan University of Technology, Isfahan, Iran}

\author{V. Salari}
\affiliation{Institute for Quantum Science and Technology, University of Calgary, Calgary, T2N 1N4, Alberta, Canada}
\affiliation{Department of Physics and Astronomy, University of Calgary, Calgary, T2N 1N4, Alberta, Canada}

\author{P. Plawiak}
\affiliation{Department of Computer Science, Cracow University of Technology, Poland}

\begin{abstract}
\noindent{The process of translating a quantum algorithm into a form suitable for implementation on a quantum computing platform is crucial but yet challenging. This entails specifying quantum operations with precision, a typically intricate task. In this paper, we present an alternative approach: an automated method for synthesizing the functionality of a quantum algorithm into a quantum circuit model representation. Our methodology involves training a neural network model using diverse input-output mappings of the quantum algorithm. We demonstrate that this trained model can effectively generate a quantum circuit model equivalent to the original algorithm. Remarkably, our observations indicate that the trained model achieves near-perfect mapping of unseen inputs to their respective outputs.}
\end{abstract}

\date{\today}
\maketitle


\section{Introduction}
Quantum computing is a type of computation that uses quantum mechanical phenomena to store and process data differently from classical computers \cite{Nielsen, Sood}. In the field of quantum computing, numerous researchers endeavor to invent new quantum algorithms that can significantly outperform their classical counterparts. For the execution and implementation of these algorithms, one common model is to represent them in the form of a quantum circuit \cite{Shor, Mosca} consisting of basic quantum gates so that they can be executed on quantum computers \cite{Qsynthesis1, Qsynthesis2, Qsynthesis3, Qsynthesis4}. By decomposing a quantum algorithm into a sequence of quantum operations, one establishes a quantum computation model for the given task. This can be achieved using a quantum circuit or any other quantum computing model of interest. However, finding new quantum algorithms is very hard \cite{Mosca}, and looking for a set of quantum gates that can implement the desired algorithm can be challenging. Another more significant problem is the lack of such algorithms for many problems. For this reason, we seek a way to find the quantum circuit that implements the desired functionality of a quantum problem and performs the process of converting that specification to the quantum circuit automatically.

In addition, machine learning is well-known for performing tasks such as predicting or classifying information. Neural networks and deep learning are among the popular machine learning algorithms in recent years \cite{Mehlig}. Recently, many research studies have aimed to combine quantum computing and machine learning. Quantum machine learning \cite{Schuld2}, optimization \cite{Mariam}, and quantum-inspired machine learning algorithms \cite{Moore, Hakemi} are among the most investigated research areas.

In this paper, we propose an approach for building a given quantum circuit using neural networks. Many similarities exist between the steps involved in making computations in a quantum system and the way fully connected neural networks perform the training to find the best parameters or weights. Therefore, after training the neural network, it can be mapped to the desired quantum circuit.

Quantum evolution \cite{Morley} is defined as applying a unitary operation $U$ to the current quantum state of the system. A unitary operation, which is the operation of the evolution of a quantum state, is defined by a unitary matrix. From linear algebra, an invertible complex square matrix, $U$, is unitary if $UU^{\dagger}=U^{\dagger}U=I$, where $U^{\dagger}$ is the transpose conjugate of $U$, and $I$ is the identity matrix. Therefore, we require that our neural network have unitary weights to map to a quantum circuit easily.
This paper aims to use neural networks to decompose quantum computation into basic quantum gates and to optimize it. In quantum computation, this task is sometimes known as quantum logic synthesis. We propose a way to construct and keep unitary weight matrices required for the quantum computations and algorithms, and these weight matrices are used in training the neural network model. 

\section{Quantum circuit synthesis with neural networks}

To produce a quantum circuit as the output model of a quantum synthesis algorithm, we can consider two representations to specify and display the quantum circuit.

In the first representation- using basic quantum gates- a series of basic quantum gates are considered, and according to their characteristics, they are used to produce the desired circuit. Based on the specific approach, the space of different configurations is explored and searched. Works like \cite{Davis} use this method for synthesising the quantum circuit. 
In another representation, the concept of a unitary matrix is applied. According to the definition \cite{Ma, Quintino}, a quantum circuit can be represented with a product of unitary matrices. In this way, if we multiply the state of the input qubits by that matrix, the state of the output qubits will be obtained. If the number of input qubits of the circuit is $N$, the unitary matrix of that circuit will be a $2^N \times 2^N$ matrix, and the state vector has the dimension $2^n \times 1$. This method is suitable for the process we are considering. We will seek to find the desired matrix based on unitarity, and if we can find the matrix correctly, this matrix will be converted into the desired quantum circuit. This final synthesis action is called transpile. It is not the focus of this work, and it is the next stage in the quantum synthesis process. We have used existing approaches for this stage. Another related work \cite{Camps} uses block encoding for approximate quantum circuit synthesis. These block encodings are embeddings of an operator in a larger unitary matrix. 

In one of the works in quantum circuit synthesis \cite{Wu}, authors convert a quantum circuit block into a unitary matrix and then use a synthesis tool to convert it to another quantum circuit block, which is optimised to reduce the number of CNOTs and algorithm time.

In one of the most recent studies, Fürrutter et.al introduced a method denoising diffusion models (DMs) for quantum circuit synthesis \cite{Furrutter}, where quantum circuits are encoded into three-dimensional tensors. The model is trained to iteratively denoise these encoded circuits, guided by text-conditioning, to generate gate-based quantum circuits, demonstrating high accuracy in tasks like entanglement generation and unitary compilation.

\begin{figure}
    \centering
    \includegraphics[scale=0.24]{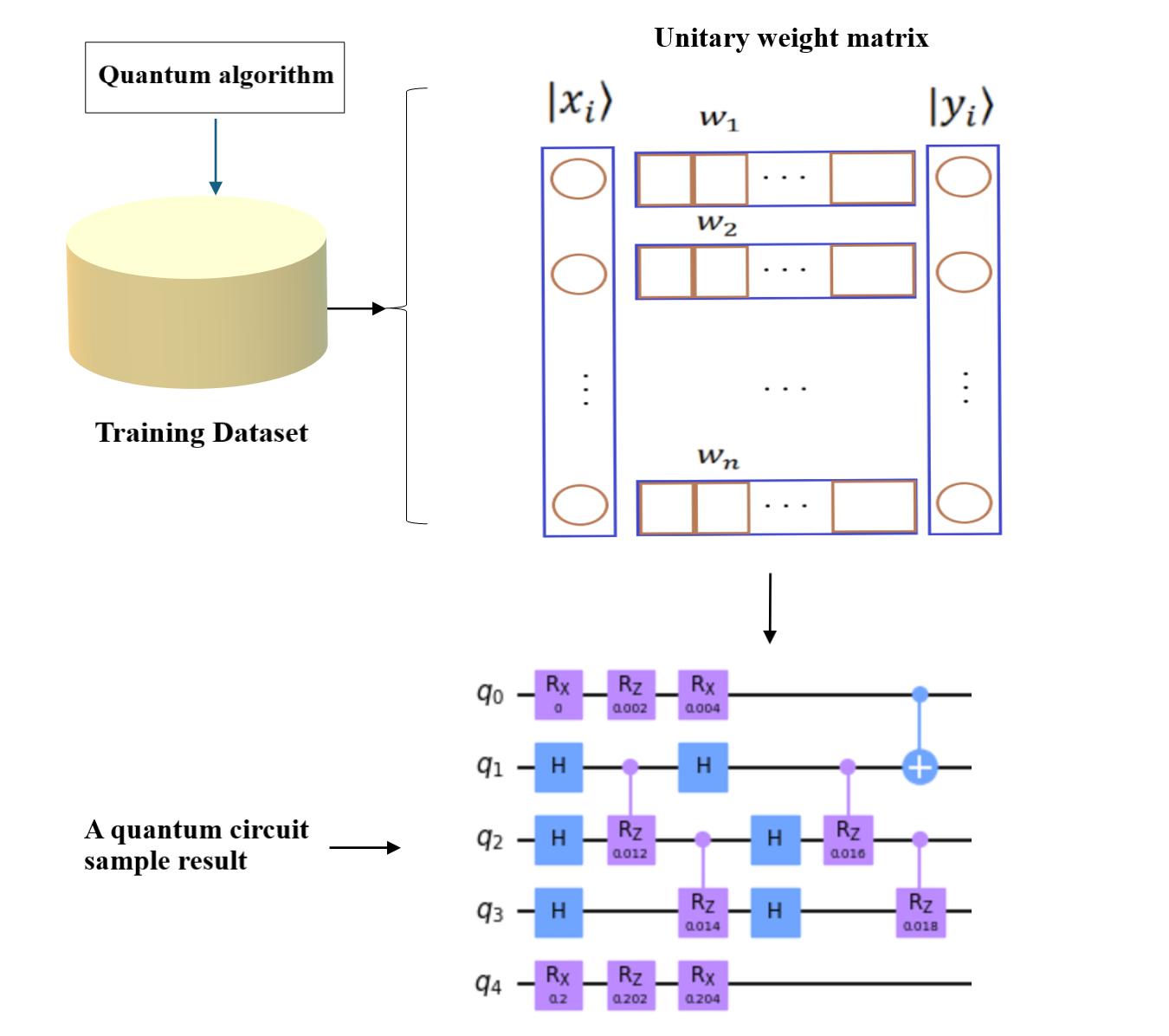}
    \caption{General protocol for quantum circuit synthesis including neural network training and the resulting quantum circuit after the transpilation in Qiskit.}
    \label{fig:enter-label}
\end{figure}

Different methods can be used to search and find the desired unitary matrix. The method we are investigating is the use of neural networks. To obtain the output state of qubits of a quantum circuit, it is enough to multiply their input state in the unitary matrix of the desired circuit. This is exactly done in multi-layer perceptron (MLP) neural networks. Since a neural network can learn the appropriate weights to solve a problem by using a data set, consider the unitary matrix elements of the circuit as the weights of a single-layer MLP neural network. We can find circuit unitary matrix elements by training the neural network using a suitable data set. Therefore, for this purpose, we use a single-layer MLP neural network. 

In the input of a quantum circuit, we have an $n$-qubit state, which is represented as a vector $\Psi = \Sigma_{i}c_{i}|\psi_{i}\rangle$. This quantum register is multiplied by the unitary matrix of the circuit. The quantum register has a dimension of $2^N\times1$. Therefore, if the number of qubits equals $N$, the desired neural network must have $2^N$ inputs.

To provide the architecture of a quantum neural network for building a quantum circuit, we need to present the framework of the quantum computation task as the data to train the neural network model. Quantum computation (QC) is accomplished by a series of basic quantum operations known as quantum gates in the quantum circuit model, which act on qubits.
A complex vector describes a qubit in $C^2$ and usually is represented in bra-ket notation as $|\psi> = \alpha |0> + \beta |1>$. Let $n$ be the number of qubits involved in a given quantum computation. Similarly, an $n$ qubit system is represented as a vector in $C^n$.

We encode input and output data for a quantum circuit as vectors and train the model using a simple neural network to obtain unitary weight parameters for synthesising the quantum circuit.

A quantum circuit is a model used to represent a given quantum algorithm or quantum computation in general as a series of quantum gates coming from left to right while applied on qubits. The way we build a quantum circuit that implements a given algorithm or a quantum computation is known as the synthesis of quantum circuits or quantum logic synthesis.
We mentioned earlier the most related work in quantum circuit synthesis, and there are many more approaches in the literature for synthesising a quantum circuit starting from different levels of abstraction. This work uses neural networks to build a quantum circuit from a given quantum computation, such as a quantum algorithm. We propose a method to make the weights of neural networks as unitary operations. Unitary weights correspond to any individual time step of quantum computation in a quantum circuit model and must be unitary.

\section{Neural networks for building quantum circuits}

Due to the similarity in the architecture of a quantum circuit and a neural network, they can be trained to build a model for a quantum computing ansatz.
The neural network model motivates our method for building the quantum circuit ansatz to find the appropriate weight parameters. Until now, in the works concerning machine learning algorithms for quantum circuits, neural networks have not been used to build the entire time evolution of the quantum computation using input-output mappings of the quantum algorithms.

In training neural networks with unitarity constraint, two approaches have been mainly used to solve this challenge, mostly in unitary recurrent neural networks, where the main aim is to solve the vanishing and exploding gradients.

The first method considers an initial weight matrix composed of several parameterised unitary matrices \cite{Arjovsky}. Then, the parameters in the unitary matrices are learned through backpropagation without the need for expensive computations after weight updates.
The second method is to carry out the entire training process without unitarity constraint and, at the end of each iteration, convert the weights matrix to a unitary matrix using the existing algorithms. This algorithm maps the total space of $N \times N$ matrices to the space of unitary $N \times N$ matrices \cite{Kiani}.
The advantage of this method is that the entire network training process is the same; no changes are made, and all the tools and algorithms implemented in the frameworks are used. The drawback is the computational time required to calculate the unitary matrix at each iteration and after each weight update.

In this work, the operation of the quantum computation is described by a unitary matrix $U(\theta)$, and the idea is to build this unitary using the training process of a neural network algorithm.

The set of unitary matrices denoted by $U(n)$ is a subgroup of general linear groups $GL(n,\mathbb{C})$, and it is also a matrix Lie group \cite{Hall, Kiani, Kiani2}.

\begin{figure}[hbt!]
\centering
\includegraphics[scale=0.28]{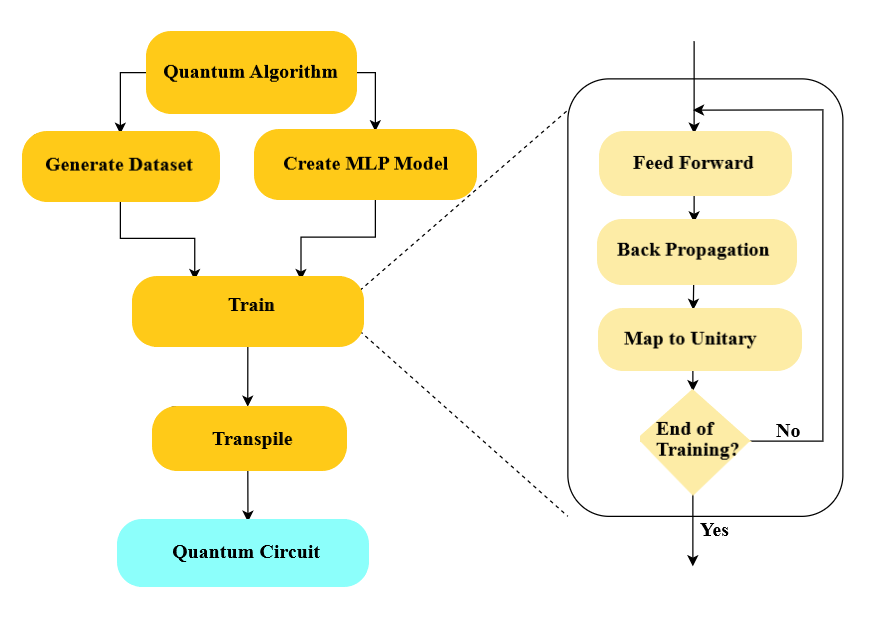}
\caption{The flowchart summary of the entire process of designing quantum circuits by using neural networks.}
\label{all-process}
\end{figure}

The ansatz of the quantum circuit is not accomplished by finding only the angles of the rotations like in the usual works of quantum neural networks instead, the whole ansatz is built based on the neural network model and decomposition approach. So, this work aims at creating and optimising the whole unitary operation of the quantum computation and not just optimizing the parameters of unitary operations in the quantum circuits. At the same time, it is not devoted to building a quantum neural network model but to using neural networks for building a general quantum circuit model.
The classical optimiser, our neural network model, is responsible for building the overall unitary $U$ of the quantum circuit and the final ansatz QC.
Figure \ref{all-process} represents a summary of the entire process.

The training of the desired neural network to find the unitary matrix of the quantum circuit is different from the usual training of a neural network. The main challenge in training this network is that the weight matrix of the network must be unitary; if not, the learned weights of the neural network do not meet the unitary condition of quantum computation.


\section*{Training neural networks for quantum circuit design}
Putting these two parts together, this work aims to use unitary weight neural networks to train the input-output mapping of quantum computation.

It is proved that if $U$ is unitary, then the columns and rows of $U$ form an orthonormal basis of $C^n$.

In this work, the matrix $W$ is a unitary matrix, but with weight updates using the above formula, $W'$ is no longer a unitary matrix. We apply the Gram-Schmidt process at each iteration to convert $W'$ into a unitary matrix.

After the network training, we use the transpile function to convert the final unitary matrices into quantum circuits. The transpile function in quantum computing is used to convert a given quantum circuit into an equivalent circuit that can be executed on a specific hardware or simulator or to optimize the quantum circuit. It considers the constraints and limitations of the target device and optimizes the circuit accordingly.

When converting unitary matrices into quantum circuits, the transpile function helps decompose the unitary matrix into a sequence of elementary gates supported by the target device. This process is known as gate synthesis or gate decomposition.

The transpile function analyses the gates required to implement the unitary matrix and checks if they are available on the target device. If not, it tries to find an alternative set of gates that can be used to approximate the desired transformation as closely as possible.

Additionally, transpilation also considers other factors, such as gate connectivity constraints, gate error rates, and gate duration on the target device. It rearranges and optimizes gates to minimize errors and improve performance.

Transpiling unitary matrices into quantum circuits using the transpile function ensures that the resulting circuit is compatible with a specific hardware or simulator, considering its limitations and optimizing for better execution.

\subsection{Training Process}
For the input vector \textbf{$x$}; here a quantum state \textbf{$|x\rangle$}, and an initial weight matrix \textbf{$U\in U_n(C)$}, the set of $n$-by-$n$ unitary matrices, we perform the training process in neural networks.
We initialize the unitary weight matrix \( U_{n\times{n}} \). We use the same initialization as \cite{Helfrich} where all entries of the unitary matrix are zero except for $2 \times 2$ submatrices $u_i$ along the diagonal as follows:

\[
U = \begin{bmatrix}
u_1 & & \\
& \ddots & \\
& & u_{\lfloor n/2 \rfloor}
\end{bmatrix}
\]

where each block matrix $u_i$ is defined as follows:

\[
u_i = \begin{bmatrix}
0 & v_i \\
-v_i & 0
\end{bmatrix}
\]

and \( v_i = \sqrt{\frac{1 - \cos(t_i)}{1 + \cos(t_i)}} \) and \( t_i \) is is drawn uniformly from the interval \([0, \frac{\pi}{2}]\).

Then we calculate the outputs (\( O \)) and then the gradients (\( G \)) of the loss function $L$ with respect to the unitary weight matrix \( U \).
After gradient calculation, Gram-Schmidt is applied to the updated weight matrix to make a unitary weight matrix.
\begin{center}
$U' =$ Gram-Schmidt $(W')$.
\end{center}


Given a set of vectors $v_1, v_2, ..., v_n$ in an $n$-dimensional vector space,
the Gram-Schmidt process proceeds as follows:

\begin{itemize}
    \item 
    Initialisation:
    $u_1=\dfrac{v_1} { ||v_1|| }$
    
    \item 
    Orthogonalisation (Orthonormalisation): 
    Orthogonalise (or Orthonormalise) $v_k$ for $ k=2,...,n $ with respect to previous vectors:
    
    For $k=2$, $v_2' = v_2 - (v_2 \cdot u_1)u_1$, so 
    $ (u_1 \cdot v_2')=0 $ and in the same way we do orthogonalization.
    
    \item 
    Normalisation:
    Normalise $V_k$ (only for orthonormal basis):
    $$u_k = \dfrac{v_k'} { ||v_k'|| } $$
    
    \item 
    Finalisation:
    After the process, the resulting set $u_1, u_2, ..., u_n$ forms the orthogonal (or orthonormal) basis.
    
\end{itemize}
If the vectors are linearly dependent, the Gram-Schmidt still produces an orthogonal set, but the vector $x_k$ is zero for the least $k$ that $x_k$ is a linear combination of $x_1, x_2, \dots, x_{k-1}$ \cite{Horn}. That is because if
$(u_k \cdot v_{k+1}')=0$ we conclude:
\begin{itemize}
\item 
$u_k$ and $v_{k+1}'$ are orthogonal, or
\item
$v_{k+1}'$ is zero.
\end{itemize}

The following statements are equivalent for a given $M \in \mathbb{C}^{n\times{n}}$:

The rows of $M$ and the columns of $M$ exhibit a linearly independent set.

$M$ is a nonsingular matrix.

\begin{theorem}[\cite{Horn}] 
    For any matrix \( M \in \mathbb{C}^{n\times{n}} \), 
    there is a small value \( \delta > 0 \) 
    such that \( M + \epsilon I \) is nonsingular 
    for \( \epsilon \in \mathbb{C} \) and \( 0 < |\epsilon| < \delta \).
\end{theorem}

We consider this remark in the algorithm such that if we get a zero column in the matrix, it goes back a step. By adding a small amount to one of the components of the vector $V_{k+1}$, the vectors $V_{k+1}$ and $u_k$  would be linearly independent. For more details, see the following Proposition \ref{prop}:

\begin{prop}\label{prop}.
Let we have a set of vectors ${x_1, \cdots, x_n}$, and $x_1' = \dfrac{x_1} { ||x_1|| }$. For $i=2, \cdots, n$ put 
$x_n'=x_n-\text{proj}(x_n,x_1')-\cdots-\text{proj}(x_n,x_{n-1}')$, so
$(x_n' \cdot x_i')=0$, for $i=1, \cdots, n$. We conclude that $x_i' \perp x_{i-1}'$ or $x_i'=0$, in this case, the vector $x_i'$ is dependent on previous vectors. Let $x_i=(x_{i1}, \cdots, x_{in})$, by adding a partial mount to one of the non-zero components of the vector $x_{i}$, the vectors $x_i$ and $x_{i-1}'$  would be linearly independent.

Proof. Let $\epsilon, x_{i1} \neq 0$, 
$x_i=\alpha _1 x_{1}'+\cdots+ \alpha _{i-1} x_{i-1}'$
$$x_i'' = (x_{i1}+\epsilon, x_{i2}, \cdots, x_{in})$$
we claim the vectors $x_1', \cdots, x_{i-1}', x_i''$
are independent.
Proof by contradiction
$$x_i''=\beta _1 x_{1}'+\cdots+ \beta _{i-1} x_{i-1}'$$
so
$$x_{i1}=\alpha _1 x_{11}'+\cdots+ \alpha _{i-1} x_{(i-1)1}'$$
and
$$x_{i1}+\epsilon=\beta _1 x_{11}'+\cdots+ \beta _{i-1} x_{(i-1)1}'$$
so
$\epsilon = (\beta_1 - \alpha _1) x_{11}'+\cdots+ (\beta_{i-1}-\alpha _{i-1}) x_{(i-1)1}'$. In other side
$$x_{i2}=\alpha _1 x_{11}'+\cdots+ \alpha _{i-1} x_{(i-1)1}' = \beta _1 x_{11}'+\cdots+ \beta _{i-1} x_{(i-1)1}'$$
and the vectors $x_1', \cdots, x_{i-1}'$ are linear independent, therefore $$\alpha_j - \beta_j = 0, ~~~j=1,\cdots, i-1$$
this implies that $\epsilon=0$, and this is a contradiction.

\end{prop}


\section{Results and discussion}
This work considers a conventional feed-forward neural network (FNN) consisting of one fully connected layer. Having a dataset D of input-output mappings of a given quantum computation, we train the network using the unitary weight matrix $U$. The network learns the input-output mapping of the quantum computation to find the best quantum circuit for the QC.

\begin{figure}[hbt!]
    \centering
    \includegraphics[scale=0.28]{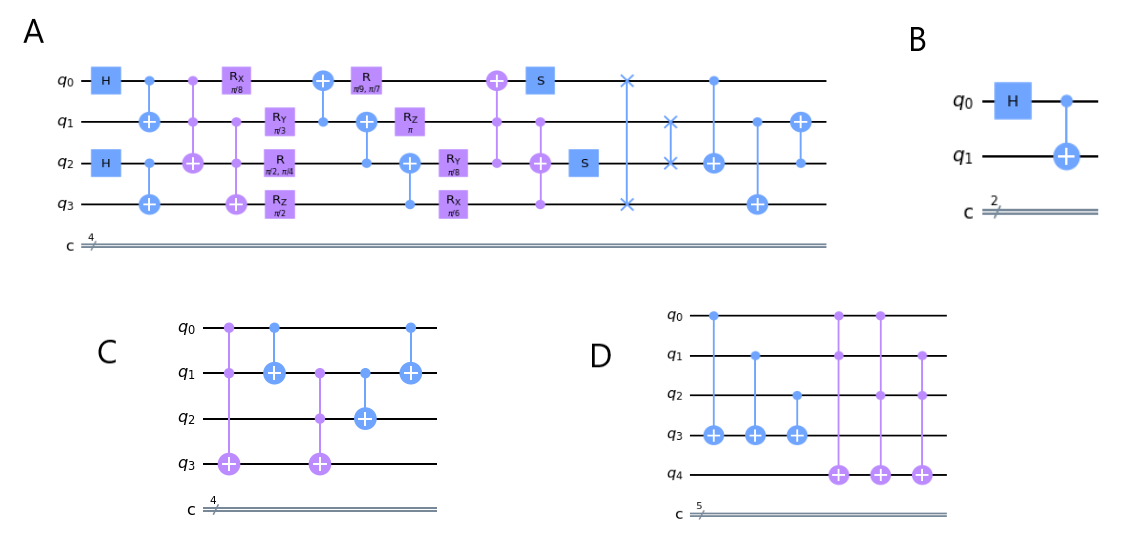}
    \caption{A) Random circuit with 4-qubits and circuit level 17. B) Implementation of entanglement state with 2-qubits and circuit level two. C) A quantum full adder with 4 qubits and circuit level five. D) A quantum full adder with 5 qubits and 6 circuit level.  }
    \label{fig:enter-label}
\end{figure}

Circuit 1 is a random circuit with four qubits and a depth 17 (Fig. 3A).
The next circuit represents the entanglement circuit with two qubits and depth two (Fig. 3B).
Circuit number 3 is a quantum full adder with four qubits and depth 5 (Fig. 3C), and finally, circuit 4 is another full adder circuit with five qubits and depth 6 (Fig. 3D).

\begin{figure}[hbt!]
    \centering
    \hspace*{-0.2cm}
    \includegraphics[scale=0.28]{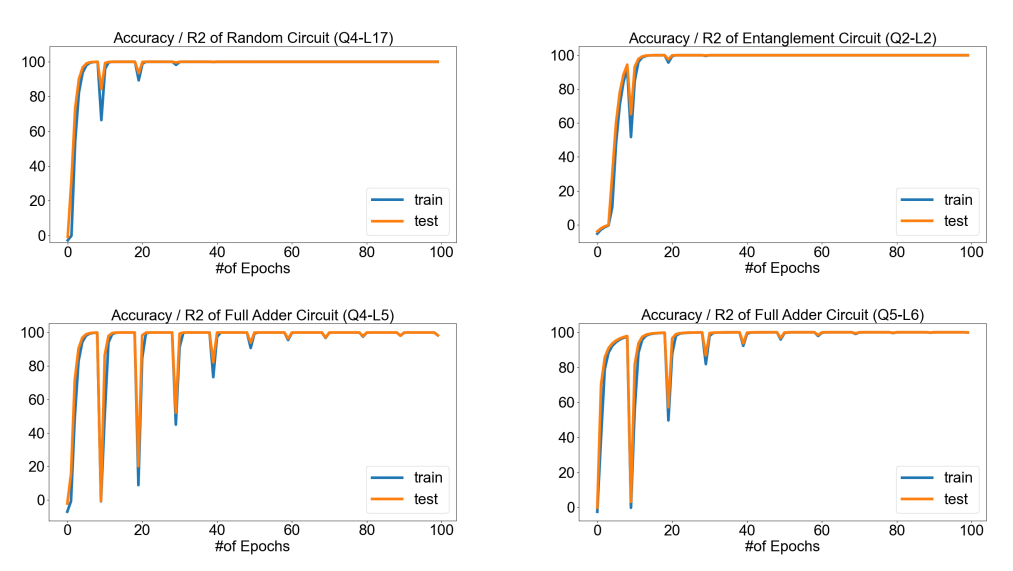}
    \caption{The results of model training (Accuracy/R2)}
    \label{accuracy}
\end{figure}

\begin{figure}[hbt!]
    \centering
    \hspace*{-0.2cm}
    \includegraphics[scale=0.28]{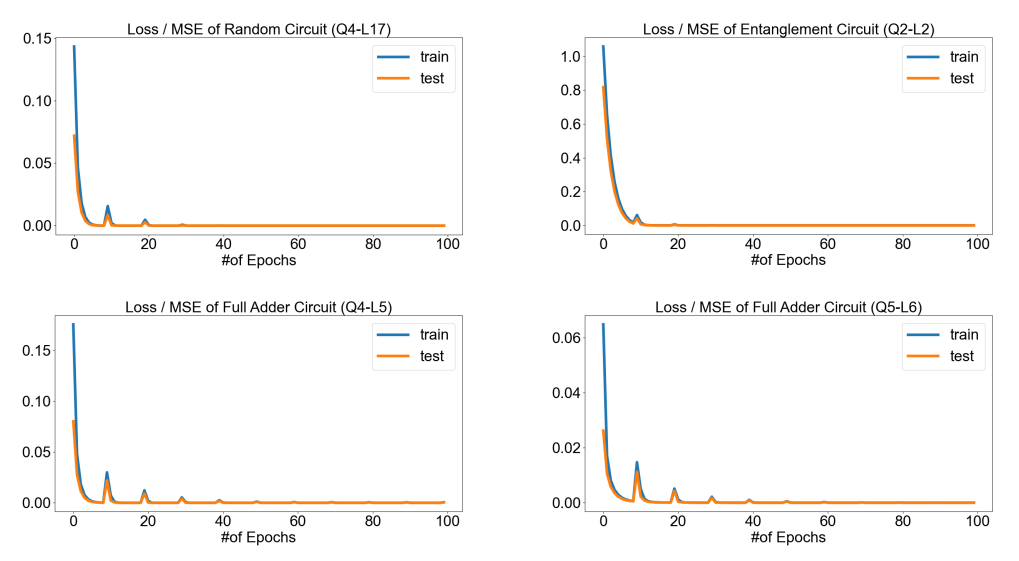}
    \caption{The results of model training (Loss/MSE)}
\label{loss}
    \label{fig:enter-label}
\end{figure}

Figure \ref{accuracy} and \ref{loss} represent training results, and Table \ref{total_results} represents the results for different quantum computation mappings consisting of random, entanglement, and two adders. The results indicate the trained model achieves near-perfect mapping of unseen inputs to their respective outputs.

\begin{table}[h!]
    \centering
    \caption{Results for different quantum circuits}
     \hspace*{-0.2cm}
    \includegraphics[scale=0.47]{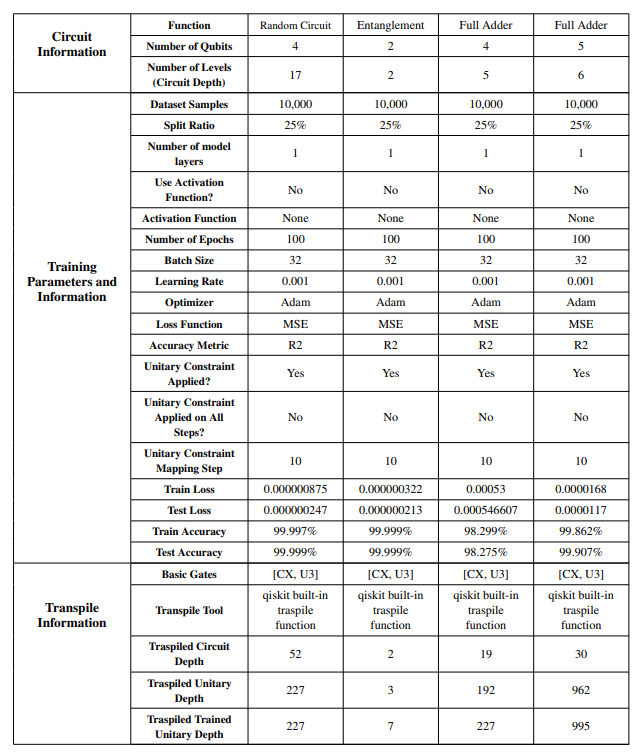}
    \label{total_results}
\end{table}


In summary, we have proposed a method to learn the unitary quantum evolution of a quantum algorithm through the use of unitary weight neural networks. Our experimental results show that the obtained model is generated almost perfectly to unseen data.
One can also extend this approach to train a multi-layer neural network and build a multi-level quantum circuit from scratch.

\section{Data Availability}
To train the neural network, it is required to feed it with a sufficient number of samples. 
As a dataset, our model needs the input-output mapping of quantum computation which means for each input quantum state what output state should be produced by the underlying quantum computation model which in this work is a quantum circuit.
All data are available from the below link:
\url{https://github.com/hossein-aminii/Quantum-Circuit-NN}





\noindent\textbf{Acknowledgments:}
\noindent V. S. is very thankful for NSERC and NRC grants of Canada. \\



\noindent\textbf{Correspondence} and requests for materials should be addressed to M.Z. (zomorodi@outlook.com). 




\section{Appendix}


Research at the intersection of neural networks and quantum computing spans several areas. It is categorised into quantum-inspired algorithms for neural networks \cite{Menneer}, neural network-inspired quantum computing \cite{Murakami}, and the design and implementation of quantum algorithms for solving neural network tasks on quantum computers \cite{Schuld}. Furthermore, several studies focused on simulating quantum computations using neural networks on classical computers \cite{Jonsson}.

In some investigations, quantum computation and quantum circuit models collaborate with classical neural networks to achieve specific objectives. For example, in \cite{Mina} and \cite{Mina2}, quantum circuits for natural language processing are integrated into classical neural network models, employing techniques such as long short-term memory (LSTM) for translation tasks.

In the field of quantum circuit synthesis and optimization, some researchers concentrate on decomposing the unitary operations of quantum algorithms into a universal set of low-dimensional and physically realizable quantum gates \cite{Matteo}.
In several other researches, the decomposition of the quantum functionality into a set of basic quantum gates has been explored \cite{Zomorodi2, Davis2}. Some works consider a high-level synthesis approach, starting from the very beginning of a problem specification in a quantum description language \cite{Zomorodi3, Liu2}.

In our study, we trained various models using different optimisation techniques, primarily stochastic gradient descent. Our work aligns with the first two categories mentioned earlier. While we leverage neural networks to construct quantum computing ansatz, our approach involves quantum-inspired computations within the neural network architecture.

Additionally, research in quantum circuit synthesis using artificial intelligence techniques can be classified into two main categories: evolutionary algorithms for evolving quantum circuits \cite{Sasamal, Krylov, Mukherjee}, and machine learning methods applied to quantum circuit synthesis \cite{Murakami}. It is worth noting that numerous mathematical and heuristic approaches also contribute to building quantum gate-level systems \cite{Saeedi}.

The utilisation of neural networks for the automated synthesis of quantum circuits has been explored in \cite{Murakami}. The methodology involves the random generation of quantum circuits followed by training a neural network to predict the probability of selecting quantum gates. However, the neural network in their study was not directly responsible for generating quantum circuits but rather for prioritising the generation probability of quantum gates.

In another study \cite{1Kiani}, the learning of unitary transformations within the context of quantum computation and deep learning was investigated. The authors compared unitary matrix parameterisation in both quantum and classical computation, suggesting the potential for leveraging unitary parameterisations to design quantum architectures with deep learning algorithms.

When neural networks are employed for constructing a quantum circuit, the input to the neural network consists of complex numbers. Two approaches exist for designing neural networks with complex-valued inputs. The first approach treats complex values' real and imaginary parts as two separate real-valued inputs \cite{Trabelsi}. The second approach involves designing neural networks to handle complex-valued inputs directly \cite{Wang}. In this work, we adopt the first approach.

Complex-valued neural networks (CVNNs) are neural networks with complex numbers as input and network parameters, and the network outputs can be either real or complex numbers based on the application. In a neural network with the state of a quantum system as input, a CVNN is utilised with complex numbers in the input, network parameters, and outputs.

On the other hand, using unitary matrices as weights of neural networks has been explored in several studies, including significant work in \cite{Arjovsky1}. The authors investigated unitary recurrent neural networks with unitary weight matrices in the hidden layers, employing complex-valued matrices to implement this unitary weight model.

\subsection{Unitary weight matrix}
We assume weight matrices of the neural network as $2^n\times 2^n$ matrices over the field $\textbf{C}$ composed of complex numbers. The set of all of these matrices is denoted by $M_n(C)$ and abbreviated to $M_n$. A matrix $U$ in $M_n$ is unitary if $U^*U=I$, where $U^\dagger$ is the conjugate transpose of $U$ and $I$ is the identity matrix.  

As the dimension of input and output is the same in quantum computations, therefore having $n$ input neurons and $n$ output neurons, the weight matrix has dimension $n$$\times$$n$.
\begin{equation}
    W_{n{\times}n}{\times}I_{n{\times}1}=O_{n{\times}1}
\end{equation}
Unitary weight matrix in neural networks has applications in other fields than quantum circuit synthesis. Recently, this area has had good results in recurrent neural networks as well as convolutional neural networks.
The unitary matrix $U$ of quantum computation has some properties that make it great as the parameters of a neural network. It preserves the norm, length and angle between vectors in the rows and columns of the matrix. These properties reduce the vanishing and exploding gradients of neural networks in general \cite{Bengio}.
A unitary matrix $U$ is a complex matrix whose inverse equals its conjugate transpose, i.e. $UU^T=U^TU=I$ or $U^{(-1)}=U^T$. The unitary matrix has orthonormal columns and rows. It means that the rows of a unitary matrix form an orthogonal set; this is also true for the columns of the unitary matrix.
By having unitary matrices as the weights of a neural network, we preserve the strength of the outputs without the need for normalization of the layer. We also produce unitary operations for the quantum operations in each time step.

The quantum circuit model is a very popular and basic quantum computing model. In this model, qubits are represented as horizontal lines, and quantum gates are vertically connected to these lines using different predefined symbols. The computation is performed from left to right, and each set of parallel computations is considered a time step or time slot.

The information flow in this circuit model is the same as the information flow in a neural network model, where the input is connected to different units in the first layer, which processes the inputs. Then, it flows to the next layer, repeated in a few layers until we have the final output. Due to the similarity between these two models of computations, we propose a decomposition of a quantum computation into a quantum circuit based on a neural network.

\subsection{Training neural network for quantum circuit design}
Putting these two parts together, this work aims to use unitary weight neural networks to train the input-output mapping of quantum computation.

It is proved that if $U$ is unitary, then the columns and rows of $U$ form an orthonormal basis of $C^n$.

\section*{Optimization in Classical Neural Networks}

The neural network maps the inputs to the outputs in layer $l$ according to the following forward-pass formula:

\begin{center}
$ O^l=f(W^l x^l) $
\end{center}

Here, $O$ is the output of the layer, $f$ is the activation function, $W$ is the weight matrix, which is a unitary weight in our problem, and $x$ is the input to the particular layer.

\[ O = f(U_{n\times{n}} x) \]
\[ L = g(O(W_{n\times{n}})) \]
\[ G = \frac{\partial L}{\partial U} \]
Input $x$ with dimension $2^n$ is the quantum state and input to a quantum computation. The state of $x$ is defined as a vector $|x\rangle$ in $2^n$ dimensional Hilbert space $\mathcal{H}^{\otimes{n}}$ which is the feature space in our problem \cite{Maria}.
A training set $\mathcal{T}= \{|X\rangle, |Y\rangle\}$ of inputs and outputs is given to the algorithm and each vector in the input or output set is a vector in the $\mathcal{H}^{\otimes{n}}$ Hilbert space.

Weight updates are accomplished using the following update rule:
\[ W' = U - \eta G \], where \( \eta \) is the learning rate and set to a small number between 0 and 1.

\textbf{}

In classical neural networks, optimisation involves minimising the error or loss function on a given dataset. 
Optimisation aims to train the neural network parameters to perform well on the task it's designed for.
Unitary weight neural networks are considered as an optimization problem under unitarity constraint. More precisely, the problem is defined as:
optimising a loss function $\mathcal{L}: \mathbb{C}^{n\times n} \rightarrow \mathcal{R}$ subject to unitarity constraint defined as $UU^\dagger = U^{\dagger}U = I$. The parameter space is $n\times n$ unitary matrices belonging to the Lie group of unitary operations $U(n)$. This problem can be changed to an unconstraint problem in the Lie group of unitary matrices $U(n)$.

Back-propagation is the basic algorithm to optimise loss function $L$ and is used to compute the gradients $(\nabla L)$ of the loss function with respect to the weights and biases of the neural network. mean squared error (MSE) is used for the loss function. These gradients indicate how the loss changes with small variations in the parameters.

Gradient descent is applied to update and optimise the weights:

$$ W' = W - \eta \frac{\partial{L}}{\partial{W}} $$
This is the classical approach for updating the weights in a neural network, where $\eta$ is the learning rate and $\frac{\partial{L}}{\partial{W}}$ is the gradient of the cost function $L$ concerning the weight matrix $W$.

After each weight update, the new weight matrix may no longer be unitary.

In this work, the matrix $W$ is unitary $U$. We use the transpile function to convert unitary matrices into quantum circuits. The transpile function in quantum computing is used to convert a given quantum circuit into an equivalent circuit that can be executed on a specific hardware or simulator or to optimise the quantum circuit. It considers the constraints and limitations of the target device and optimises the circuit accordingly.

When converting unitary matrices into quantum circuits, the transpile function helps decompose the unitary matrix into a sequence of elementary gates supported by the target device. This process is known as gate synthesis or gate decomposition.

The transpile function analyses the gates required to implement the unitary matrix and checks if they are available on the target device. If not, it tries to find an alternative set of gates that can be used to approximate the desired transformation as closely as possible.

Additionally, transpilation also considers other factors such as gate connectivity constraints, gate error rates, and gate duration on the target device. It rearranges and optimises gates in order to minimise errors and improve performance.

Transpiling unitary matrices into quantum circuits using the transpile function ensures that the resulting circuit is compatible with a specific hardware or simulator, taking into account its limitations and optimising for better execution.

\subsection{Gram-schmidt process for unitary Weight matrix}

Apply the Gram-Schmidt process to orthogonalise vectors in the layer's weight matrices. Normalise the column vectors to obtain the final unitary weight matrix \( W_{nn} \):\\
\[ \text{Given } W_{nn} = [w_1, w_2, \ldots, w_n],\]\\
 apply Gram-Schmidt to obtain orthonormal set \[[u_1, u_2, \ldots, u_n] \]
\[ \text{Normalize each column vector: } u_k = \frac{u_k}{\|u_k\|} \]

\subsection{Training Process}
Assuming an initial unitary weight matrix \( U_{n\times{n}} \), we calculate the outputs (\( O \)) and then the gradients (\( G \)) of the loss function $L$ with respect to the unitary weight matrix \( U \):
\[ O = f(U_{n\times{n}} x) \]
\[ L = g(O(W_{n\times{n}})) \]
\[ G = \frac{\partial L}{\partial U} \]
Input $x$ with dimension $2^n$ is the quantum state and input to a quantum computation. The state of $x$ is defined as a vector $|x\rangle$ in $2^n$ dimensional Hilbert space $\mathcal{H}^{\otimes{n}}$ which is the feature space in our problem \cite{Maria2}.
A training set $\mathcal{T}= \{|X\rangle, |Y\rangle\}$ of inputs and outputs is given to the algorithm and each vector in the input or output set is a vector in the $\mathcal{H}^{\otimes{n}}$ Hilbert space.

Weight updates are accomplished using the following update rule:
\[ U' = U - \eta G \] where \( \eta \) is the learning rate and is set to a small number.

It is important to mention that the Gram-Schmidt process may lead to numerical instability when dealing with ill-conditioned or nearly linearly dependent sets of vectors. Robust algorithms like Modified Gram-Schmidt or Householder transformations may be considered.

\subsection{Algorithms}

The algorithms of the training process in the normal state and the state where we have the limitation of the matrix being unitary are in the form of algorithms 1 and 2 as follows. As can be seen, Algorithms 1 and 2 are similar and differ only in one process. In Algorithm 2, after updating the weights, we map the weights matrix to the space of unitary matrices.
The process of mapping to the unitary space, if done after each iteration, may slow down the process. Hence, mapping can be done only once after several iterations. However, it should be noted that mapping must be done in the last iteration of the training. Algorithm 3 shows this issue.\\

\begin{widetext}

\begin{algorithm}[H]
\caption{The normal training process in supervised learning}
\begin{algorithmic}[1]
\STATE \textbf{model}: MLP(hyperparameters) \textcolor{blue}{\textit{   // initialize with proper hyperparameters}}\
\STATE \textbf{initialize\_weights()} \textcolor{blue}{\textit{     // initialize model with random weights}}
\STATE \textbf{W}: model.weights
\STATE \textbf{dataset}: load\_dataset()
\FOR{iteration \textbf{from} 1 \textbf{to} epochs}
    \FOR{input\_batch \textbf{in} dataset.next\_batch()}
        \STATE Predicted: model(input\_batch) \textcolor{blue}{\textit{     // feed forward}}
        \STATE Loss: calculate\_loss(true\_labels, Predicted)
        \STATE Update\_Through\_BP(Loss, W) \textcolor{blue}{\textit{   // weights update through BackPropagation}}
    \ENDFOR
\ENDFOR
\STATE \textbf{Done.}
\end{algorithmic}
\end{algorithm}
\begin{algorithm}[H]
\caption{training with \textbf{Unitary} constraint}
\begin{algorithmic}[1]
   \STATE \textbf{model}: MLP(hyperparameters) \textcolor{blue}{\textit{    // initialize with proper hyperparameters}}
    \STATE \textbf{initialize\_weights()} \textcolor{blue}{\textit{     // initialize model with random weights}}\
    \STATE \textbf{W}: model.weights\;
    \STATE \textbf{dataset}: load\_dataset()\;
    \FOR{ iteration \textbf{from} 1 \textbf{to} epochs}
        \FOR{ input\_batch \textbf{in} dataset.next\_batch()}
           \STATE Predicted: model(input\_batch) \textcolor{blue}{\textit{  // feed forward}}\   
            \STATE Loss: calculate\_loss(true\_labels, Predicted)\;
            \STATE Update\_Through\_BP(Loss, W)\textcolor{blue}{\textit{    // weights update through BackPropagation}}\
            \STATE \textbf{Map\_Weights\_To\_Unitary\_Space(W)} \;
        \ENDFOR
    \ENDFOR
    \STATE \textbf{Done.}
    \end{algorithmic}
    \end{algorithm}

\begin{algorithm}[H]
\caption{training with \textbf{Unitary} constraint and mapping step to improve performance}
\begin{algorithmic}[1]
    \STATE \textbf{model}: MLP(hyperparameters) \textcolor{blue}{\textit{   // initialize with proper hyperparameters}}\
    \STATE \textbf{initialize\_weights()} \textcolor{blue}{\textit{     // initialize model with random weights}}\
    \STATE \textbf{W}: model.weights\;
    \STATE \textbf{dataset}: load\_dataset()\;
    \STATE \textbf{define MS}\textcolor{blue}{\textit{      // Mapping Step}}\
    \FOR{ iteration \textbf{from} 1 \textbf{to} epochs}
        \FOR{ input\_batch \textbf{in} dataset.next\_batch()}
            \STATE Predicted: model(input\_batch) \textcolor{blue}{\textit{     // feed forward}}\   
            \STATE Loss: calculate\_loss(true\_labels, Predicted)\;
            \STATE Update\_Through\_BP(Loss, W)\textcolor{blue}{\textit{    // weights update through BackPropagation}}\
            \STATE \textbf{Map\_Weights\_To\_Unitary\_Space(W)} \
        \ENDFOR
    \ENDFOR
    \STATE \textbf{Done.}\
    \end{algorithmic}
    \end{algorithm}
\end{widetext}


    
	
    

	


\end{document}